\documentclass[twocolumn,showpacs,preprintnumbers,nofootinbib,amsmath,amssymb]{revtex4}
\usepackage{epsfig}
\usepackage{amssymb}

\catcode`\@=11

\newcommand{\be}[1]{\begin{equation}\label{#1}}
\newcommand{\ee}{\end{equation}}
\newcommand{\ba}[1]{\begin{eqnarray}\label{#1}}
\newcommand{\ea}{\end{eqnarray}}
\newcommand{\rf}[1]{(\ref{#1})}
\newcommand{\nn}{\nonumber}

\newcommand{\f}{\mathrm{f}}

\newcommand{\const}{\mbox{\rm const}\,}

\begin{document}
\title{Extra dimensions and Lorentz invariance violation}

\author{Viktor Baukh}
\email{bauch_vGR@ukr.net} \affiliation{Department of Theoretical
Physics and Astronomical Observatory, Odessa National University,
2 Dvoryanskaya St., Odessa 65026, Ukraine}
\author{Alexander Zhuk}
\email{zhuk@paco.net} \affiliation{Department of Theoretical
Physics and Astronomical Observatory, Odessa National University,
2 Dvoryanskaya St., Odessa 65026, Ukraine}
\author{Tina Kahniashvili}
\email{tinatin@phys.ksu.edu} \affiliation{ CCPP, New York
University, 4 Washington Place, New York, NY 10003, USA \\
National Abastumani Astrophysical Observatory, 2A Kazbegi Ave,
Tbilisi, GE-0160 Georgia}

%
%
%
%

%
\begin{abstract} We consider effective model where photons interact
with scalar field corresponding to conformal excitations of the
internal space (geometrical moduli/gravexcitons). We demonstrate
that this interaction results in a modified dispersion relation
for photons, and consequently, the photon group velocity depends
on the energy implying the propagation time delay effect. We
suggest to use the experimental bounds of the time delay of gamma
ray bursts (GRBs) photons propagation as an additional constrain
for the gravexciton parameters.
\end{abstract}

 \maketitle

\vspace{.5cm}


Lorentz invariance (LI) of physical laws is one of the corner
stone of modern physics. There is a number of experiments
confirming this symmetry at energies we can approach now. For
example, on a classical level, the rotation invariance has been
tested in Michelson-Morley experiments, and the boost invariance
has been tested in Kennedy-Torhndike experiments \cite{reviews}.
 Although, up to now, LI is well established experimentally, we
cannot say surely that at higher energies it is still valid.
Moreover, modern astrophysical and cosmological data (e.g. UHECR,
dark matter, dark energy, etc) indicate for a possible LI
violation (LV). To resolve these challenges, there are number of
attempts to create
 new physical models, such as M/string theory, Kaluza-Klein
models, brane-world models, etc. \cite{reviews}.

In this paper we investigate LV test related to  photon
dispersion measure (PhDM). This test is based on the LV effect of
a phenomenological energy-dependent speed of photon
\cite{a98,ellis00,km01,sarkar,myers,piran03,jlm03}, for recent
studies see Ref. \cite{MP06} and references therein.

The formalism that we use is based on the analogy with
electromagnetic waves propagation in a magnetized medium, and
extends previous works \cite{cfj90,jlm03,Tina1}. In our model,
instead of propagation in a magnetized medium, the
electromagnetic waves are propagating in vacuum filled with a
scalar field $\psi$. LV occurs because of an interaction term
$\f(\psi )F^2$ where $F$ is an amplitude of the electromagnetic
field. Such an interaction might have different origins. In the
string theory $\psi $ could be a dilaton field
\cite{string,damour-pol}. The field $\psi$ could be  associated
with geometrical moduli. In brane-world models the similar term
describes an interaction between the bulk dilaton and the Standard
Model fields on the brane \cite{zhuk}. In Ref. \cite{klp}, such
an interaction was obtained in $N=4$ super-gravity in four
dimensions. In Kaluza-Klein models the term $\f(\psi )F^2$ has
the pure geometrical origin,  and it appears in the effective,
dimensionally reduced, four dimensional action (see e.g.
\cite{kubyshin,GSZ}). In particular, in reduced
Einstein-Yang-Mills theories, the function $f(\psi )$ coincides
(up to a numerical prefactor) with the volume of the internal
space. Phenomenological (exactly solvable) models with spherical
symmetries were considered in Refs.~\cite{melnikov}. To be more
specific, we consider the model which is based on the reduced
Einstein-Yang-Mills theory \cite{GSZ}, where the term $\propto
\psi F^2$ describes the interaction between the conformal
excitations of the internal space (gravexcitons) and photons. It
is clear that the similar LV effect exists for all types of
interactions of the form $\f(\psi )F^2$ mentioned above.

Obviously, the interaction term $\f(\psi )F^2$ modifies the
Maxwell equations, and, consequently, results in a modified
dispersion relation for photons. We show that this modification
has rather specific form. For example, we demonstrate that
refractive indices for the left and right circularly polarized
waves coincide with each other. Thus, rotational invariance is
preserved.  However, the speed of the electromagnetic wave's
propagation in vacuum differs from the speed of light $c$. This
difference implies the time delay effect which can be measured
via high-energy GRB photons propagation over cosmological
distances (see e.g. Ref. \cite{MP06}). It is clear that
gravexcitons should not overclose the Universe and should not
result in variations of the fine structure constant. These
demands lead to a certain constrains for gravexcitons (see Refs.
\cite{GSZ,iwara}). We use the time delay effect, caused by the
interaction between photons and gravexcitons, to get additional
bounds on the parameters of gravexcitons.

The starting point of our investigation is the Abelian part of
D-dimensional action of the Einstein-Yang-Mills theory:
\be{1.0}
S_{EM}= -\frac12 \int\limits_{M}d^{D}x\sqrt{|g|}\, F_{MN}F^{MN}\, ,
\ee
where the D-dimensional metric, $g = g_{MN}(X)dX^M\otimes dX^N =
g^{(0)}(x)_{\mu \nu}dx^{\mu}\otimes dx^{\nu} + a_1^2(x)g^{(1)}$,
is defined on the product manifold $M=M_0\times M_1$. Here, $M_0$
is the $(D_0=d_0+1)$-dimensional external space. The
$d_1$-dimensional internal space $M_1$ has a  constant curvature
with the scale factor $a_1(x) \equiv L_{Pl}\exp \beta^1(x)$.
Dimensional reduction of the action \rf{1.0} results in the
following effective $D_0$-dimensional action \cite{GSZ}
\be{1.00} \bar S_{EM}= -\frac {1}{2}
\int_{M_0}d^{D_0}x\sqrt{|\tilde g^{(0)}|} \left[\left(1 -
\mathcal D \kappa_0 \psi \right) F_{\mu\nu}F^{\mu \nu}\right], \ee
which is written in the Einstein frame with the $D_0$-dimensional
metric, $\tilde g^{(0)}_{\mu \nu}=(\exp d_1\bar
\beta^{1})^{-2/(D_0-2)}g^{(0)}_{\mu \nu}$. Here, $\kappa_0 \psi
\equiv -\bar \beta^1 \sqrt{(D_0-2)/d_1(D-2)}\ll 1$ and $\bar
\beta^1 \equiv \beta^1 - \beta^1_0$ are  small fluctuations of
the internal space scale factor over the stable background
$\beta^1_0$ ($0$ subscript denotes the present day value). These
internal space scale-factor small fluctuations/oscillations have
the form of a scalar field (so called gravexciton \cite{GZ1}) with
a mass $m_{\psi}$ defined by the curvature of the effective
potential (see for detail \cite{GZ1}). Action \rf{1.00} is defined
under the approximation $\kappa_0 \psi < 1$ that obviously holds
for the condition\footnote{In the brane-world model the prefactor
$\kappa_0$ in the expression for $\kappa_0\psi$ is replaced by the
parameter proportional to $M^{-1}_{EW}$ \cite{zhuk}. Thus, the
smallness condition holds for $\psi < M_{EW}$.} $\psi < M_{Pl}$.
$\kappa_0^2 = 8\pi/M^2_{Pl}$ is four dimensional gravitational
constant, $M_{Pl}$ is the Plank mass, $\mathcal D = 2
\sqrt{d_1/[(D_0-1)(D-1)]}$ is a model dependent constant. The
Lagrangian density for the scalar field $\psi$ reads:
$\mathcal{L}_{\psi} = \sqrt{|\tilde g^{(0)}|}(-\tilde
g^{\mu\nu}\psi_{,\mu}\psi_{,\nu}-m_{\psi}^2 \psi \psi)/2$. For
simplicity we assume that $\tilde g^{0}$ is the flat
Friedman-Lemaitre-Robertson-Walker (FLRW) metric with the scale
factor $a(t)$.

Let's consider Eq. \rf{1.00}. It is worth of noting that the
$D_0$-dimensional field strength tensor, $F_{\mu \nu}$, is gauge
invariant.\footnote{Eq. \rf{1.00}
 can be rewritten in the more familiar form $\bar
S_{EM}=-(1/2)\int_{M_0}d^{D_0}x\sqrt{|\tilde g^{(0)}|} \bar
F_{\mu\nu}\bar F^{\mu \nu}$ \cite{GSZ}. The field strength tensor
$\bar F_{\mu\nu}$ is not gauge invariant here.} Secondly, action
\rf{1.00} is conformally invariant in the case when $D_0=4$. The
transform to the Einstein frame does not break  gauge invariance
of the action \rf{1.00}, and the electromagnetic field is
antisymmetric as usual,
$F_{\mu\nu}=\partial_{\mu}A_{\nu}-\partial_{\nu}A_{\mu}$. Varying
\rf{1.00} with respect to the electromagnetic vector potential,
\be{1.3}
\partial_{\nu}\left[ \sqrt{-g}
\left(1-\mathcal D\kappa_0 \psi \right)F^{\mu \nu} \right]=0. \ee
The second term in the round brackets $\mathcal D\kappa_0 \psi
F^{\mu\nu}$ reflects the interaction between photons and the
scalar field $\psi$, and as we show below, it is responsible  for
LV. In particular, coupling between photons and the scalar field
$\psi$ makes  the speed of photons  different from the standard
speed of light. Eq. (\ref{1.3}) together with  Bianchi identity
(which is preserved in the considered model due to
gauge-invariance of the tensor, $F_{\mu\nu}$ \cite{GSZ}) defines
a complete set of the generalized Maxwell equations. As we noted,
action \rf{1.00} is conformally invariant in the $4D$ dimensional
space-time. So, it is convenient to present the flat FLRW metric
$\tilde g^{0}$ in the conformally flat form: $\tilde
g^{0}_{\mu\nu}=a^2\eta_{\mu \nu}$, where $\eta_{\mu\nu}$ is the
Minkowski metric.

Using the standard definition of the electromagnetic field tensor,
$F_{\mu \nu}$, we obtain  the complete set of the Maxwell
equations in vacuum,

\begin{eqnarray}
{\bf \nabla}  \cdot {\bf B}&=&0\, , \label{divB}
\\
{\bf \nabla}  \cdot {\bf E}&=& \frac{{\mathcal D}\kappa_0}{1-
{\mathcal D}\kappa_0 \psi} ({\bf \nabla} \psi \cdot {\bf E})\, ,
\label{divE}
\\
{\bf \nabla}  \times {\bf B}&=&\frac{\partial {\bf E}}{\partial
\eta}
 - \frac{{\mathcal D} {\kappa_0\dot\psi}}{1- {\mathcal
D}\kappa_0 \psi} {\bf E} \nn \\ &+& \frac{{\mathcal D}\kappa_0}{1-
{\mathcal D}\kappa_0 \psi} [{\bf \nabla} \psi \times {\bf B}]\, ,
\label{rotB}
\\
{\bf \nabla}  \times {\bf E}&=&-\frac{\partial {\bf B}}{\partial
\eta}\, , \label{rotE}
\end{eqnarray}
where all operations are performed in the Minkowski space-time,
$\eta$ denotes conformal time related to physical time $t$ as $dt
= a(\eta)d\eta$, and an overdot represents a derivative with
respect to conformal time $\eta$.

Eqs. \rf{divB} and \rf{rotE} correspond to Bianchi identity, and
since it is preserved, Eqs. \rf{divB} and \rf{rotE} keep their
usual forms. Eqs. \rf{divE} and \rf{rotB} are modified due to
interactions between photons and gravexcitons ($\propto \mathcal
\kappa_0 \psi$). These modifications have
 simple physical meaning: the interaction between photons and
the scalar field $\psi$ acts as an effective electric charge
$e_{eff}$. This effective charge is proportional to the scalar
product of the $\psi$ field gradient and the ${\bf E}$ field, and
it vanishes for an homogeneous $\psi$ field. The modification of
Eq. (\ref{rotB}) corresponds to an effective current ${\bf
J}_{eff}$, which depends on both electric and magnetic fields.
This effective current is determined by variations of the $\psi$
field over the time ($\dot\psi$) and space ($\nabla \psi$). For
the case of a homogeneous $\psi$ field the effective current is
still present and LV takes place. The modified Maxwell equations
are conformally invariant.
  To account for the expansion of the Universe we
  rescale the field components as${\bf{B,
E}} \rightarrow {\bf{B, E}}~ a^2$ \cite{grasso}.

To obtain a dispersion relation for photons, we use the Fourier
transform between position and wavenumber spaces as,
\begin{eqnarray}
   {\bf F}({\bf k}, \omega) &=&
\int \int d \eta ~  d^3\!x \,
   e^{-i(\omega \eta - {\bf k} \cdot {\bf x})} {\bf F}({\bf x}, \eta )\, ,
\nonumber
\\
{\bf F}({\bf x}, \eta ) &=& \frac{1}{(2\pi)^4}\int \int {d\omega}~
{d^3\!k}
   e^{i(\omega \eta -  {\bf k}\cdot {\bf x})}
{\bf F}({\bf k}, \omega )\, . \label{field1}
\end{eqnarray}
Here, ${\bf F}$ is a vector function describing either the
electric or the magnetic field, $\omega$ is the angular frequency
of the electro-magnetic wave measured today, and ${\bf k}$ is the
wave-vector. We assume that the field $\psi$ is an oscillatory
field with the frequency $\omega_{\psi}$ and the momentum $\bf q$,
so $ \psi({\bf x}, \eta )=Ce^{i(\omega_\psi \eta - {\bf q} \cdot
{\bf x})}\, ,\quad C=\const$.
 Eq. (\ref{divB}) implies ${\bf B} \perp
{\bf k}$. Without loosing of generality, and for simplicity of
description we assume that the wave-vector ${\bf k}$ is oriented
along the ${\bf z}$ axis. Using Eq. \rf{rotE} we get ${\bf E}
\perp {\bf B}$.

A linearly polarized wave can be expressed as a superposition of
 left (L, $-$)
and right (R, $+$) circularly polarized (LCP and RCP) waves. Using
the polarization basis of Sec.~1.1.3 of
Ref.~\cite{varshalovich89}, we derive $E^\pm =(E_x \pm i
E_y)/\sqrt{2}$. Rewriting  Eqs. (\ref{divB}) - \rf{rotE} in the
components,{\footnote{We have defined the system of 6 equations
with respect to 6 components of the vectors ${\bf E}$ and ${\bf
B}$. This system has non-trivial solutions only if its determinant
is nonzero. From this condition we get the dispersion relation.
The Faraday rotation effect is absent if the matrix has a diagonal
form. }} for LCP and RCP waves we get,
\begin{equation}
(1-n_{+}^2)E^{+}=0, ~~~~~~~~~~~(1-n_{(-)}^2)E^{-}=0\, ,
\label{dispersion1}
\end{equation}
where $n_{+}$ and $n_-$ are refractive indices for RCP and LCP
electromagnetic waves
\begin{equation}
n_+^2= \frac{k^2  \left[1-{\mathcal D} \kappa_0 \psi
(1+q_{z}/k)\right]}{\omega^2\left[1-{\mathcal D} \kappa_0\psi(1+
\omega_\psi /\omega)\right]}= n_-^2\, . \label{n-}
\end{equation}
 In the case when LI is preserved the electromagnetic waves propagating
in vacuum have $n_+=n_-=n=k/\omega \equiv 1$. For the
electromagnetic waves propagating in the magnetized plasma,
$k/\omega \neq 1$, and the difference between the LCP and RCP
refractive indices describes the Faraday rotation effect, $\alpha
\propto \omega (n_+ - n_-)$ \cite{krall}. In the considered model,
since $n_+=n_-$ the rotation effect is absent, but the speed of
electromagnetic waves propagation in vacuum differs from the speed
of light $c$ (see also Ref. \cite{Cantcheff:2004dn} for LV induced
by electromagnetic field coupling to other generic field). This
difference implies the propagation time delay effect, $\Delta t =
\Delta l (1-\partial k/\partial \omega)$ ($\Delta l$ is a
propagation distance), $\Delta t$ is the difference between the
photon travel time and that for a "photon" which travels at the
speed of light $c$. Here, $t$ is physical synchronous time. This
formula does not take into account the evolution of the Universe.
However, it is easy to show that the effect of the Universe
expansion is negligibly small.

Solving the dispersion relation as a square equation, we obtain
\be{dispersion}
\frac{\partial k}{\partial \omega}\simeq
\pm\left\{1+\frac{1}{2}\left[\frac{\omega_{\psi}^2-q_z^2}{4\omega^2}\right]({\mathcal
D} \kappa_0\psi)^2\right\}\, ,
\ee
where $\pm$ signs correspond to photons forward and backward
directions respectively.

The modified inverse group velocity \rf{dispersion} shows that the
LV effect can be measured if we know the gravexciton frequency
$\omega_{\psi}$, $z$-component of the momentum $q_z$ and its
amplitude $\psi$. For our estimates, we  assume that $\psi$ is the
oscillatory field, satisfying (in local Lorentz frame) the
dispersion relation, $\omega_\psi^2 = m_\psi^2 + {\bf q}^2$, where
$m_{\psi}$ is the mass of gravexcitons\footnote{To get physical
values of the corresponding parameters we should rescale them by
the scale factor $a$.}. Unfortunately, we do not have any
information concerning parameters of gravexcitons (some estimates
can be found in \cite{GSZ,iwara}). Thus, we intend to use possible
LV effects (supposing it is caused by interaction between photons
and gravexcitons) to set limits on gravexciton parameters. For
example, we can easily get the following estimate for the upper
limit of the amplitude of gravexciton oscillations:
\be{limit}
|\psi| \approx \frac{1}{\sqrt{\pi}\, \mathcal D}\,
\sqrt{\left|\frac{\Delta t}{\Delta l}\right|}\,
\frac{\omega}{m_{\psi}}\, M_{Pl}\, ,
\ee
where for $\omega$ and $m_{\psi}$ we can use their physical
values. In the case of GRB with $\omega \sim 10^{21}\div
10^{22}$Hz $\sim 10^{-4}\div 10^{-3}$GeV and $\Delta l \sim 3\div
5\times 10^{9}$y $\sim 10^{17}$sec the typical upper limit for the
time delay is $\Delta t \sim 10^{-4}$sec \cite{MP06}. For these
values the upper limit on gravexciton amplitude of oscillations
is{\footnote{We thank R. Lehnert to point that in addition of the
time delay effect the Cherenkov effect could be used to constrain
the electromagnetic field and $\psi$ field coupling strength
\cite{cherenkov}.}}
\be{limit2} \left|\kappa_0\psi \right| \approx
\frac{10^{-13}\mbox{GeV}}{m_{\psi}}\, . \ee
This estimate shows that our approximation $\kappa_0 \psi <1$
works for gravexciton masses $m_{\psi}>10^{-13}$GeV. Future
measurements of the time-delay effect for GRBs at frequencies
$\omega \sim 1-10$GeV would increase significantly the limit up
to $m_{\psi} > 10^{-9}$GeV.  On the other hand, Cavendish-type
experiments \cite{dvali,cavendish}) exclude fifth force particles
with masses $m_{\psi} \lesssim 1/(10^{-2}\mbox{cm}) \sim
10^{-12}$GeV which is rather close to our lower bound for $\psi$
field masses. Respectively we slightly shift the considered mass
lower limit to be $m_{\psi} \geq 10^{-12}\mbox{GeV}$.  These
masses considerably higher than the mass corresponding to the
equality between the energy densities of the matter and radiation
(matter/radiation equality), $m_{eq}\sim H_{eq}\sim 10^{-37}$GeV,
where $H_{eq}$ is the Hubble "constant" at matter/radiation
equality. It means that such $\psi$-particles start to oscillate
during the radiation dominated epoch (see appendix).
Another bound on the $\psi$-particles masses comes from the
condition of their stability. With respect to decay $\psi \to \gamma
\gamma$ the life-time of $\psi$-particles is $\tau \sim
(M_{Pl}/m_{\psi})^3t_{Pl}$ \cite{GSZ}, and the stability conditions
requires that the decay time should be greater than the age of the
Universe. According this we consider light gravexcitons with masses
$m_{\psi} \le 10^{-21} M_{Pl} \sim 10^{-2}\mbox{GeV} \sim 20 m_e$
(where $m_e$ is the electron mass).

As an additional restriction arises from the condition that such
cosmological gravexcitons should not overclose the observable
Universe. This reads $m_{\psi} \lesssim m_{eq}(M_{Pl}/\psi_{in})^4$
which implies the following restriction for the amplitude of the
initial oscillations: $\psi_{in}\lesssim
\left(m_{eq}/m_{\psi}\right)^{1/4}M_{PL} << M_{Pl}$ \cite{iwara}.
Thus, for the range of masses $10^{-12}\mbox{GeV}\leq m_{\psi}\leq
10^{-2}\mbox{GeV}$, we obtain respectively $\psi_{in}\lesssim
10^{-6}M_{Pl}$ and $\psi_{in}\lesssim 10^{-9}M_{Pl}$. According to
Eq. \rf{a.3}, we can also get the estimate
for the amplitude of oscillations of the considered gravexciton at
the present time. Together with the non-overcloseness condition,
we obtain from this expression that $|\kappa_0\psi| \sim 10^{-43}$
for $m_{\psi}\sim 10^{-12}$GeV and $\psi_{in}\sim 10^{-6}M_{Pl}$
and $|\kappa_0\psi| \sim 10^{-53}$ for $m_{\psi}\sim 10^{-2}$GeV
and $\psi_{in}\sim 10^{-9}M_{Pl}$.
Obviously, it is much less than the upper limit \rf{limit2}. Note,
as we mentioned above, gravexcitons with masses $m_{\psi}\gtrsim
10^{-2}$GeV can start to decay at the present epoch. However,
taking into account the estimate $|\kappa_0\psi| \sim 10^{-53}$,
we can easily get that their energy density $\rho_{\psi} \sim
(|\kappa_0\psi|^2/8\pi)M_{Pl}^2m_{\psi}^2\sim
10^{-55}\mbox{g}/\mbox{cm}^3$ is much less than the present energy
density of the radiation $\rho_{\gamma}\sim
10^{-34}\mbox{g}/\mbox{cm}^3$. Thus, $\rho_{\psi}$ contributes
negligibly in $\rho_{\gamma}$. Otherwise, the gravexcitons with
masses $m_{\psi} \gtrsim 10^{-2}$GeV should be observed at the
present time, which, obviously, is not the case.


Additionally, it follows from Eq. (42) in Ref. \cite{GSZ} that to
avoid the problem of the fine structure constant variation, the
amplitude of the initial oscillations should satisfy the
condition: $\psi_{in} \lesssim 10^{-5}M_{Pl}$ which, obviously,
completely agrees with our upper bound $\psi_{in} \lesssim
10^{-6}$GeV.

Summarizing we shown that LV effects can give additional
restrictions on parameters of gravexcitons. First, we found that
gravexcitons should not be lighter than $10^{-13}$GeV. It is very
close to the limit following from the fifth-force experiment.
Moreover, experiments for GRB at frequencies $\omega
> 1$GeV can result in significant shift of this lower limit
making it much stronger than the fifth-force estimates. Together
with the non-overcloseness condition, this estimate leads to the
upper limit on the amplitude of the gravexciton initial
oscillations. It should not exceed $\psi_{in}\lesssim 10^{-6}$GeV.
Thus, the bound on the initial amplitude obtained from the fine
structure constant variation is one magnitude weaker than our one
even for the limiting case of the gravexciton masses. Increasing
the mass of gravexcitons makes our limit stronger. Our estimates
for the present day amplitude of the gravexciton oscillations,
following from the obtained above limitations, show that we cannot
use the LV effect for the direct detections of the gravexcitons.
Nevertheless, the obtained bounds can be useful for astrophysical
and cosmological applications. For example, let us suppose that
gravexcitons with masses $m_{\psi}>10^{-2}$GeV are produced during
late stages of the Universe expansion in some regions and GRB
photons travel to us through these regions. Then, Eq. \rf{a.3} is
not valid for such gravexcitons having astrophysical origin and
the only upper limit on the amplitude of their oscillations (in
these regions) follows from Eq. \rf{limit2}. In the case of TeV
masses we get $|\kappa_0\psi|\sim 10^{-16}$. If GRB photons have
frequencies up to 1 TeV, $\omega \sim 1$TeV, then this estimate is
increased by 6 orders of magnitude.


\bigskip
{\bf Acknowledgments}

We thank G. Dvali, G. Gabadadze, A. Gruzinov, G. Melikidze, B.
Ratra,  and A. Starobinsky for stimulating discussions. T. K. and
A. Zh. acknowledge hospitality of Abdus Salam International Center
for Theoretical Physics (ICTP) where this work has been started.
A.Zh. would like to thank the Theory Division of CERN for their
kind hospitality during the final stage of this work. T.K.
acknowledges partial support from INTAS 061000017-9258 and
Georgian NSF ST06/4-096 grants.

\renewcommand{\theequation}{A.\arabic{equation}}

\subsection{Appendix: Dynamics of Light Gravexcitons}
\setcounter{equation}{0}

In this appendix we briefly summarize the main properties of the
light gravexcitons necessary for our investigations. The more
detail description can be found in Refs. \cite{GSZ,iwara}.

The effective equation of motion for massive cosmological
gravexciton\footnote{We have seen that the interaction between
gravexcitons and ordinary matter (in our case it is 4D-photons) is
suppressed by the Planck scale. Thus, gravexcitons are weakly
interacting massive particles (WIMPs).} is
\be{a.1}
\frac{d^2}{dt^2} \psi + (3H+\Gamma)\frac{d}{dt} \psi + m_\psi^2
\psi= 0\, ,
\ee
where $H\sim 1/t$ and $\Gamma\sim m_{\psi}^3/M_{Pl}^2$ are the
Hubble parameter and decay rate ($\psi \to \gamma \gamma$)
correspondingly. This equation shows that at times when the Hubble
parameter is less than the gravexciton mass: $H \lesssim m_{\psi}$
the scalar field begins to oscillate (i.e. time $t_{in}\sim
H^{-1}_{in}\sim 1/m_{\psi}$ roughly indicates the beginning of the
oscillations):
\be{a.2}
\psi \approx C B(t) \cos (m_{\psi}t +\delta)\, .
\ee
We consider cosmological gravexcitons with masses
$10^{-12}\mbox{GeV}\leq m_{\psi}\leq 10^{-2}$GeV. The lower bound
follows both from the fifth-force experiments and Eq.
\rf{limit2}. The upper bound follows from the demand that the
life-time of these particles (with respect to decay $\psi \to
\gamma \gamma$) is larger than the age of the Universe: $\tau =
1/\Gamma \sim \left(M_{Pl}/m_{\psi}\right)^3 t_{Pl} \ge
10^{19}\mbox{sec} > t_{univ} \sim 4\times 10^{17}$ sec. Thus, we
can neglect the decay processes for these gravexcitons.
Additionally, it can be easily seen that these particles start to
oscillate before $t_{eq}\sim H^{-1}_{eq}$ when the energy
densities of the matter and radiation become equal to each other
(matter/radiation equality). According to the present WMAP data
for the $\Lambda$CDM model it holds $H_{eq} \equiv m_{eq} \sim
10^{-56}M_{Pl} \sim 10^{-28}\mbox{eV}$. Thus, considered
particles have masses $m_{\psi}
>> m_{eq}$ and start to oscillate during the radiation
dominated stage. They will not overclose the observable Universe if
the following condition is satisfied: $m_{\psi} \lesssim
m_{eq}(M_{Pl}/\psi_{in})^4$, where $\psi_{in}$ is the amplitude of
the initial oscillations at the moment $t_{in}$ (see Eq. (18) in
Ref. \cite{iwara}).

Prefactors $C$ and $B(t)$ in Eq. \rf{a.2} for considered light
gravexcitons respectively read: $C\sim
(\psi_{in}/M_{Pl})\left(M_{Pl}/m_{\psi}\right)^{3/4}$ and $B(t) \sim
M_{Pl}\left(M_{Pl}t\right)^{-3s/2}$. Here, $s=1/2,2/3$ for
oscillations during the radiation dominated and matter dominated
stages, correspondingly. We are interested in the gravexciton
oscillations at the present time $t=t_{univ}$. In this case $s=2/3$
and for $B(t_{univ})$ we obtain: $B(t_{univ})\sim
t^{-1}_{univ}\approx 10^{-61}M_{Pl}$. Thus, the amplitude of the
light gravexciton oscillations at the present time reads:
\be{a.3}
|\kappa_0\psi| \sim 10^{-60} \frac{ \psi_{in}}{M_{Pl}}
\left(\frac{M_{Pl}}{m_{\psi}}\right)^{3/4}\, .
\ee


\end{document}